\documentclass{article}
\usepackage[a4paper, total={6in, 8in}]{geometry}
\usepackage[utf8]{inputenc}
\usepackage{amsmath}
\usepackage{amssymb}
\usepackage{caption}
\usepackage{subcaption}
\usepackage{float}
\usepackage{graphicx}
\usepackage{cancel}
\usepackage{xcolor}
 \usepackage{url}
\usepackage{tikz}
\usepackage{algorithm2e}
\usepackage[round]{natbib}
\usepackage{hyperref}

\title{A General Method for Resampling Autocorrelated Spatial Data}

\author{Rudy Arthur\\
{University of Exeter, Department of Computer Science,}\\
{Stocker Rd, Exeter EX4 4PY}\\
{E-mail:  r.arthur@exeter.ac.uk}
}

\begin{document}
 \maketitle

\abstract{
Comparing spatial data sets is a ubiquitous task in data analysis, however the presence of spatial autocorrelation means that standard estimates of variance will be wrong and tend to over-estimate the statistical significance of correlations and other observations. While there are a number of existing approaches to this problem, none are ideal, requiring detailed analytical calculations, which are hard to generalise or detailed knowledge of the data generating process, which may not be available. In this work we propose a resampling approach based on Tobler's Law. By resampling the data with fixed spatial autocorrelation, measured by Moran's I, we generate a more realistic null model. Testing on real and synthetic data, we find that, as long as the spatial autocorrelation is not too strong, this approach works just as well as if we knew the data generating process. 
}

\section{Introduction}

\begin{figure}[H]
    \centering
    \includegraphics[width=\textwidth]{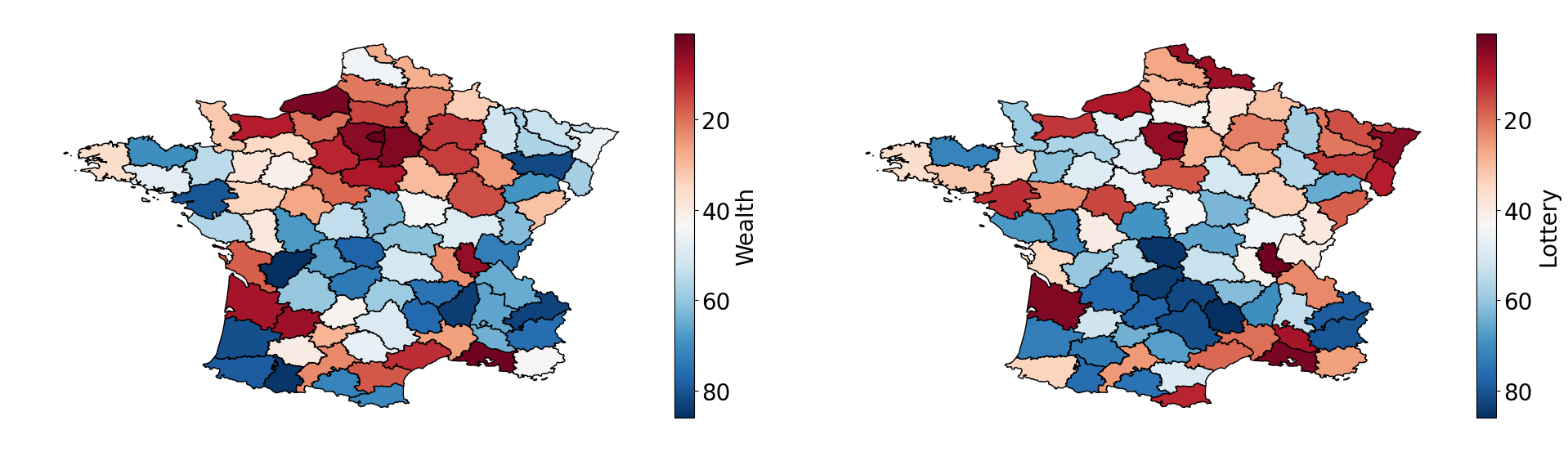}
    \caption{Left: Guerry's Wealth statistic, calculated from per capita property tax and transformed into a ranking for each sampling area (department). Right: Guerry's Lottery statistic, the per capita wager on Royal Lottery in each department, transformed into a ranking. }
    \label{fig:guerry}
\end{figure}

The problems that spatial autocorrelation causes for spatial data analysis have been well known for over a century \citep{student1914iv} and are discussed extensively in classic texts \citep{cliff1981spatial}. Indeed, spatial autocorrelation is so ubiquitous it is part of The First Law of Geography, also known as Tobler's Law \citep{tobler1970computer}, which states: \emph{everything is related to everything else, but \textbf{near things are more related than distant things}} (emphasis added). Thus, when analysing spatial data we are often confronted with patterns like Figure \ref{fig:guerry}, which plots variables from the the Guerry dataset \citep{dray2011revisit, anselin2019local}, a collection of statistics collected in France in the 1820s and 30s and aggregated at the departement level.

The Pearson correlation 
\begin{equation}\label{eqn:pearson}
    r(x, y) = \frac{\sum_{ij}^N (x_i - \bar{x})(y_i - \bar{y})}{ \sqrt{\sum_i^n (x_i - \bar{x})^2 \sum_j^N (y_i - \bar{y})^2} } 
\end{equation}
between these two data sets is $r = 0.493$. Here $x$ (and $y$) are vectors of measurements from $N$ sampling areas and $\bar{x}$ ($\bar{y}$) is the mean value across the whole region. The standard way to assess the statistical significance of Pearson correlation is to use a test based on a null model where both data sets are independent, have 0 correlation and are drawn from a bivariate normal distribution. Under these assumptions 
\begin{equation}\label{eqn:effsample}
    t = r \sqrt{\frac{N-2}{1-r^2}}
\end{equation} 
follows a t-distribution with $N-2$ degrees of freedom. From this we obtain a p-value of $1.62 \times 10^{-6}$, indicating that this level of correlation is very unlikely to arise by chance. 

However, this p-value is spuriously low, the standard significance test is not valid because of the high degree of spatial autocorrelation present in both data sets. While numerous approaches have been developed to address this issue, which we briefly review in Section \ref{sec:rev}, properly accounting for spatial autocorrelation remains a difficult issue, especially for more complicated statistics e.g. Spearman or Kendall correlation or even arbitrary real valued functions $f(x,y)$. In this work we develop a flexible method for significance testing in the presence of spatial autocorrelation, essentially by taking Tobler's Law as the null model and comparing the observed measurement against a distribution computed by resampling the data at constant autocorrelation.

In Section \ref{sec:rev} we discuss some other approaches to the problem of spatial autocorrelation. In Section \ref{sec:alg} we motivate and describe our algorithm (with some important implementation details discussed in Appendix \ref{sec:app}). In Section \ref{sec:test} we test the algorithm on synthethic, but realistic, data, showing that the correct p-values can be estimated with the method and discuss a few caveats in application. Section \ref{sec:guerry} applies the method to real data, where we find that much of the apparent correlation that would be inferred between the data using a naive approach is in fact spurious. Section \ref{sec:conc} briefly summarises our findings and discusses future work.

\section{Approaches to Spatial Autocorrelation}\label{sec:rev}

Researchers have addressed spatial autocorrelation in numerous ways, firstly, by measuring it. Standard metrics are Moran's $I$ \citep{moran1950notes} and Geary's $C$ \citep{geary1954contiguity} though there exist many others \citep{cliff1981spatial}. We will use Moran's $I$ exclusively to measure autocorrelation which will later enable a number of computational efficiencies. Given data $x$ measured at $N$ spatial sites the Moran index is defined by
\begin{equation}\label{eqn:moran}
I = \frac{N}{|W|} \frac{\sum_i \sum_j w_{ij} (x_i - \bar{x})(x_j - \bar{x})}{\sum_i (x_i - \bar{x})^2}
\end{equation}
where $|W| = \sum_{ij} w_{ij}$ and $w_{ij}$ are elements of the spatial weight matrix. The spatial weight matrix can be constructed in a number of ways. In this work we construct $w_{ij}$ such that neighbours $j$ of $i$ are assigned unit weight $\hat{w}_{ij} = 1$ in an auxillary matrix $\hat{w}$. We use the `queen' neighbour definition such that a neighbour of $i$ is another site which intersects $i$ along any line or at any point, so for example a non-edge cell on a regular grid has 8 neighbours. We set $w_{ii} = 0$ and then row normalise such that $w_{ij} = \hat{w}_{ij}/\sum_{j} \hat{w_{ij}}$. Row normalisation bounds $I$ between $-1$ and $1$. Positive values of $I$ imply positive autocorrelation (near things are more similar), values near zero imply no autocorrelation and negative values imply negative autocorrelation (near things are more different). For most data encountered in practice, remembering Tobler's Law, autocorrelation and hence $I$ is positive. For the data shown in Figure \ref{fig:guerry} the Wealth data has $I = 0.381$ and the Lottery data has $I = 0.248$.

\cite{dale2002spatial} review a number of approaches for addressing the problem spatial autocorrelation poses for statistical tests. Basic ideas like blocking or thinning can be effective but discard a lot of potentially valuable data. The most common approach is to take the view that autocorrelation reduces the amount of information contained in the data and represent this in terms of an \emph{effective} sample size $\hat{N} < N$. The effective sample size $\hat{N}$ is then used in statistical tests like equation \ref{eqn:effsample} in place of $N$. For our motivating example, Pearson correlation, the effective sample size can be calculated, under assumptions of normality and stationarity, see \cite{clifford1989assessing} and \cite{dutilleul1993modifying}. In practice this approach works well for correlation, though requires stratifying the data \citep{clifford1989assessing} and multiple estimators exist \citep{dutilleul1993modifying}. However, the major drawback is that this approach does not easily generalise to other more complicated measures of association or functions of $x$ and $y$.

Not mentioned by \cite{dale2002spatial} but also used to measure relationships between spatial data sets are alternative measures of association which incorporate the spatial weight matrix into the correlation score. Examples include Tj{\o}stheim's \citep{tjostheim1978measure}  and Wartenberg's \citep{wartenberg1985multivariate} indices, though the most accepted seems to be Lee's $L$ \citep{lee2001developing}. Defining the left spatial lag of a vector (which will be useful later) as
\begin{equation}\label{eqn:llag}
    xl_i = \sum_j w_{ij} x_j
\end{equation} and the right lag as
\begin{equation}\label{eqn:rlag}
    xr_j = \sum_i x_i w_{ij} 
\end{equation} 
\cite{lee2001developing} shows that the Moran Index can be written
\begin{equation}\label{eqn:moran2}
    I = \sqrt{ \frac{\sum_i ( xl_i - \bar{{xl}} )^2 }{\sum_i (x_i - \bar{x})^2 } } r(x, {xl})
\end{equation}
with a similar formula for the right-lagged version. This is the Pearson correlation of a variable with its spatial lag, scaled by the square root of the variance ratio of the lagged and unlagged data. Lee then defines
\begin{equation*}
    L(x,y) = \frac{\sum_i ({xl}_i - \bar{x})({yl}_i - \bar{x}) }{ \sqrt{\sum_i (x_i - \bar{x})} \sqrt{\sum_i (y_i - \bar{y})} }.
\end{equation*}
which is rewritten as
\begin{equation}
    L(x,y) = \sqrt{SSS_x} \sqrt{SSS_y} r( {xl}, {yl})
\end{equation}
Where $SSS_x$ and $SSS_y$ are `spatial smoothing scalars', an alternative measure of spatial autocorrelation for the $x$ and $y$ datasets. While Lee's $L$ effectively incorporates spatial information into a correlation score, this has the effect of making the association between \emph{identical} variables with low autocorrelation approximately equal to zero i.e. even if $r( {xl}, {yl}) = 1$, if $\sqrt{SSS_x}, \sqrt{SSS_y} \sim 0$ then $L \sim 0$, a somewhat counter-intuitive result. More importantly, as with effective sample size, while $L$ `spatialises' Pearson correlation, one would also have to spatialise other association statistics in some way, meaning that, again, the approach is hard to generalise.

The final avenue, and the one we will pursue, is a resampling approach based on simulation. The simplest approach is to permute the values $x$ and $y$, to $x'$ and $y'$, and compute the desired statistic, say $r$, using the permutations. By repeating this many times we generate a distribution of $r$ values and compare the observed value against it to estimate significance. As has been pointed out many times e.g. \citep{clifford1989assessing} and notably \citep{guillot2013dismantling}, permutation destroys  spatial autocorrelation and a naive permutation approach will not be effective for computing reliable significance scores. \cite{dale2002spatial} recommend finding a parametric representation of the underlying spatial dependence and then using Monte Carlo methods to generate samples. Below we will develop a method which, rather than model the dependence, resamples the data, such that the samples have similar autocorrelation as measured by Moran's $I$ and use these samples to perform statistical tests.

\section{Algorithm}\label{sec:alg}
\subsection{Motivating Example}\label{sec:motiv}

\begin{figure}[H]
    \centering
    \includegraphics[width=0.8\textwidth]{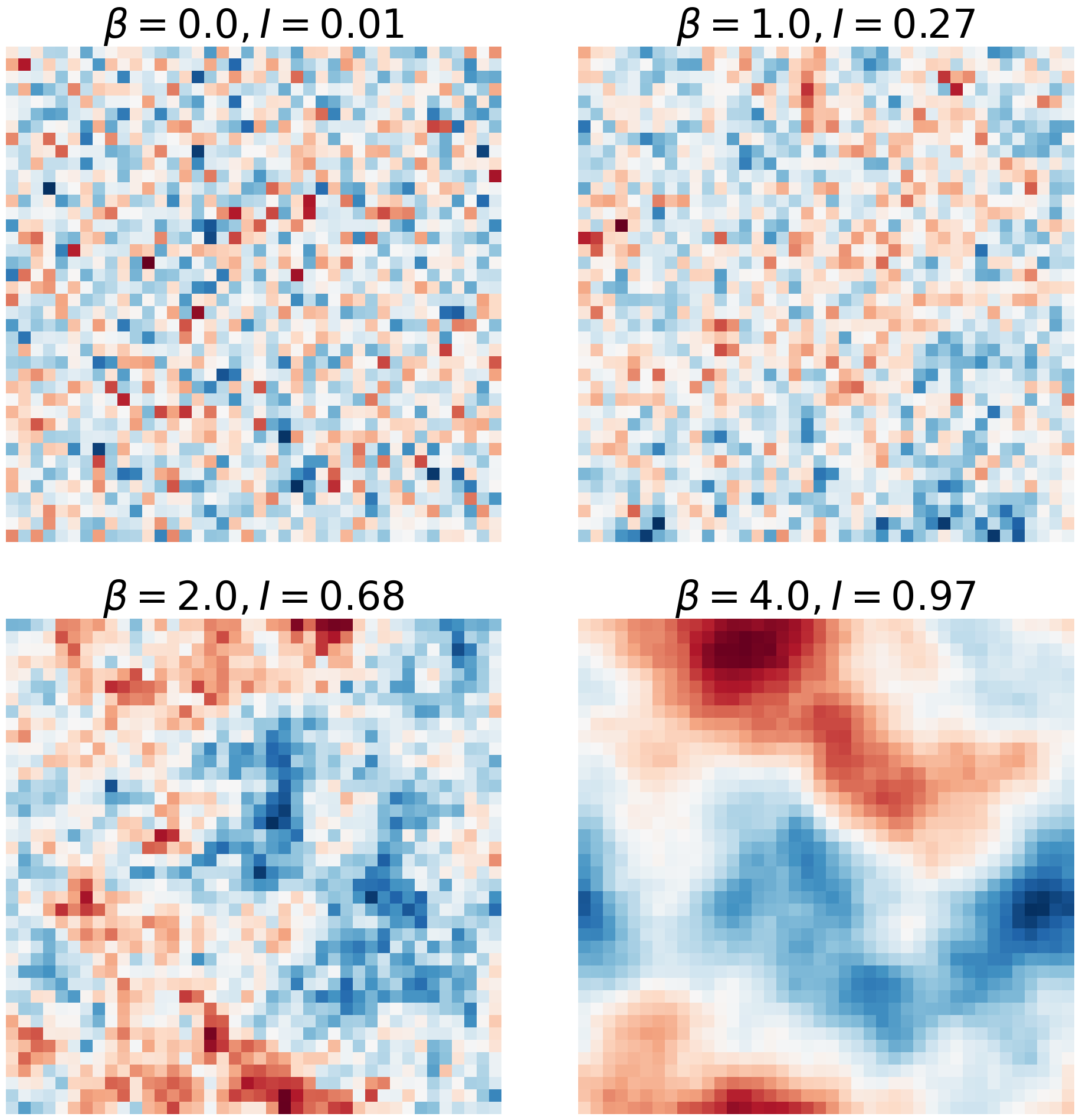}
    \caption{Random spatial data with different levels of autocorrelation, generated such that the power spectrum $S(f) \sim f^{-\beta}$. Title indicates $\beta$ value and Moran index $I$. }
    \label{fig:rfields}
\end{figure}

Following \cite{lennon2000red} we generate synthetic data with spectral density $S(f) \sim f^{-\beta}$ on a $40 \times 40$ grid. $\beta = 0$ corresponds to white noise and increasing $\beta$ increases the amount of spatial autocorrelation. Examples of some random fields generated with different $\beta$ values are shown in Figure \ref{fig:rfields}. \cite{lennon2000red} suggests (though does not seem to implement) an approach to significance testing similar to the one suggested by \cite{dale2002spatial}: generate data with the same spectral density as the observed data and use this to estimate the distribution of the test statistic. 

\begin{figure}[H]
    \centering
    \includegraphics[width=0.8\textwidth]{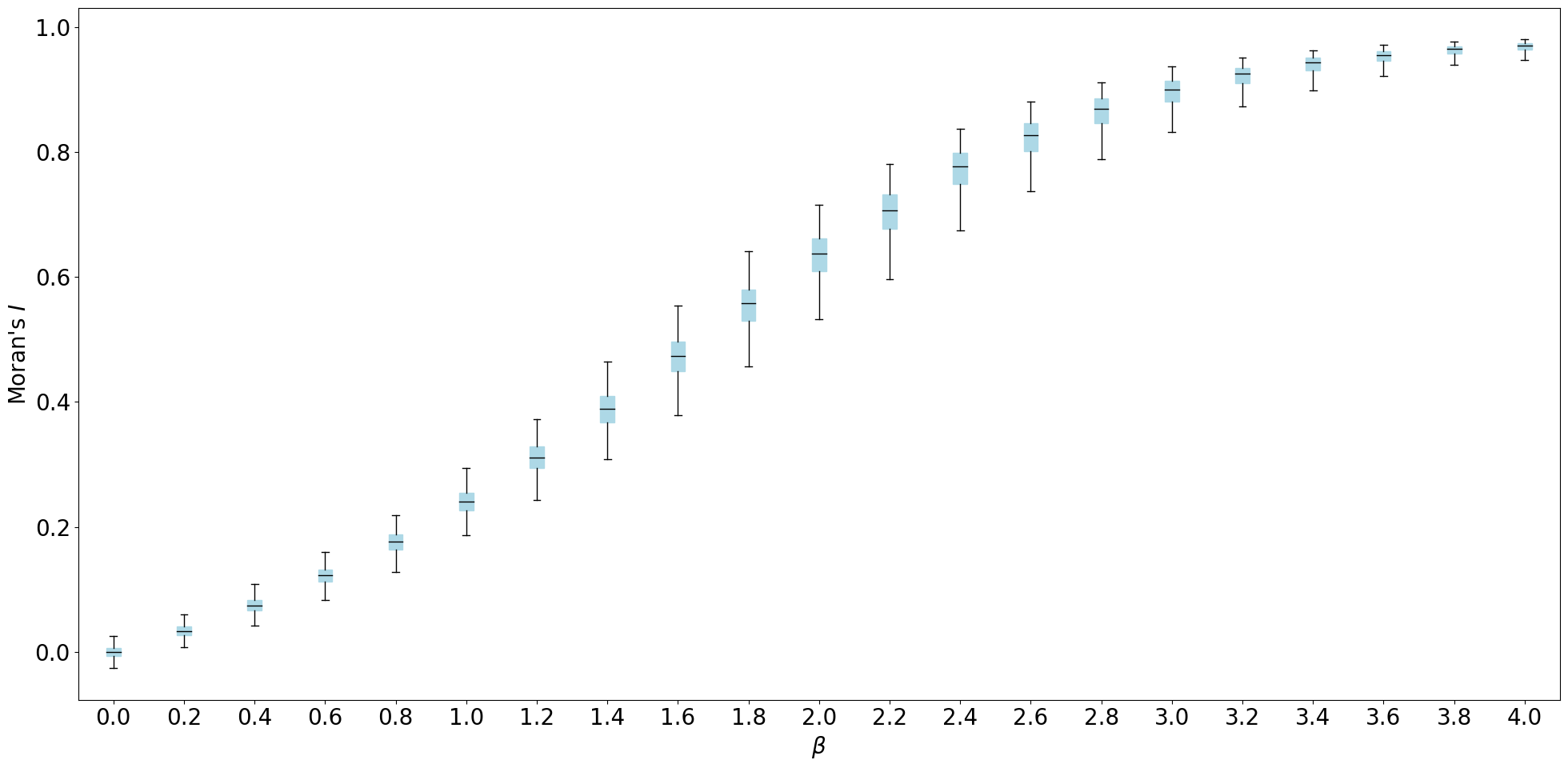}
    \caption{ Average Moran's $I$ for 1000 fields generated with different $\beta$ values on a $40 \times 40$ grid. }
    \label{fig:moranvbeta}
\end{figure}
In practice not all data sets can be described by a simple function like $S(f) = f^{-\beta}$, and even if it is the case, accurately determining $\beta$ is difficult e.g for data not sampled on a regular grid or small datasets. To overcome this, Figure \ref{fig:moranvbeta} shows the relationship between Moran's $I$ and $\beta$. When $\beta \lessapprox 2.5$ the boxes (separated by a step of 0.2 in $\beta$) don't overlap, which suggests that generating fields with similar Moran's $I$ to the observed data is equivalent to simulating new data with the same power spectrum. We do not expect a relatively simple statistic like Moran's $I$ to completely characterise all possible types of spatial autocorrelation. In Figure \ref{fig:moranvbeta} while $2.5 \lessapprox \beta$, the Moran index is saturated (very close to 1) and hence can't distinguish higher values of $\beta$. In practice values of $I \lessapprox 0.8$ are common, and here Moran's index seems to do a reasonable job characterising autocorrelation. 

\subsection{Our algorithm}

\begin{figure}[H]
    \centering
    \includegraphics[width=\textwidth]{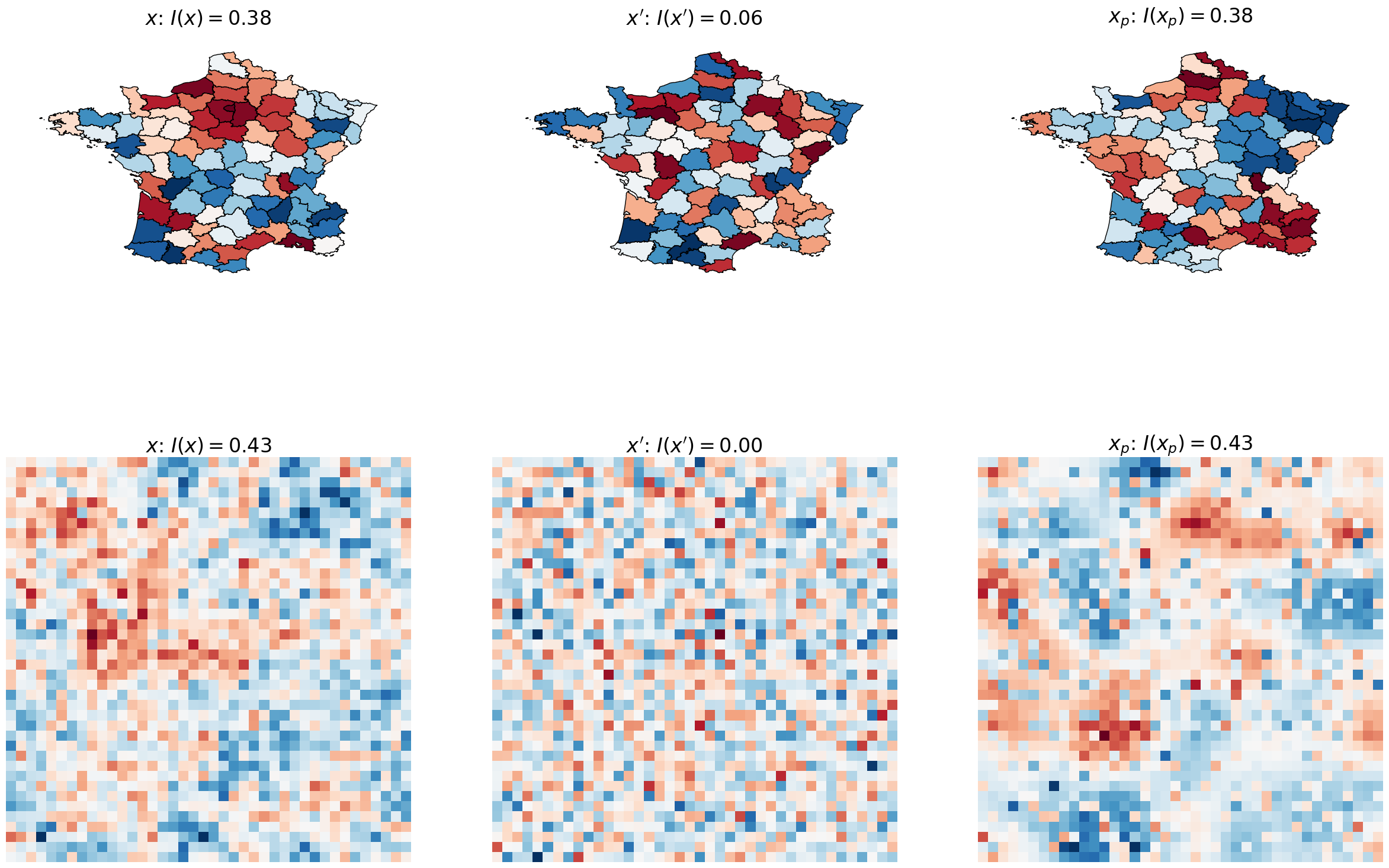}
    \caption{ Left: Original Data. Centre: Random permutation. Right: Permutation after optimising $E = (I(x) - I(x'))^2$. Top: Guerry wealth data. Bottom: Random $40 \times 40$ field with $ S(f) = f^{-1.5}$. }
    \label{fig:algexample}
\end{figure}

Motivated by the above, the algorithm we propose is conceptually quite simple. Given data $x$ with Moran index $I(x)$, resample the data by doing the following $N_{p}$ times
\begin{enumerate}
\item Randomly permute the values of $x$ to get $x'$
\item Apply a zero temperature Metropolis-Hastings (greedy) algorithm to minimise the function $E = (I(x) - I(x'))^2$ up to some target accuracy $\epsilon$, yielding a permutation $x_p$.
\end{enumerate}
We have implemented the greedy update algorithm in two ways: an approach based on swapping data values and an approach based on replacing values in $x'$ by random draws from $x$. Both yield very similar results, we refer the reader to Appendix \ref{sec:app} for the implementation details.

\section{Testing on Synthetic Data}\label{sec:test}

To evaluate the performance of the algorithm from Section \ref{sec:alg} we generate testing data following \cite{lennon2000red} as discussed in section \ref{sec:motiv}. We pick some values $\beta_x$ and $\beta_y$; generate data $x$ and $y$ with power spectra $S(f) \sim f^{-\beta_1}, f^{-\beta_2}$; compute a number of correlation statistics and evaluate their significance by creating an ensemble of $N_p$ data sets, $x^{(p)}$ and $y^{(p)}$. We use three different methods to create a testing distribution: 
\begin{enumerate}
\item Random permutations.
\item Generating new fields with $S(f) = f^{-\beta_1}$ and $S(f) = f^{-\beta_2}$.
\item The algorithm of Section \ref{sec:alg}.
\end{enumerate}
Based on the discussion above, we expect 1 to perform very badly, random permutation will destroy the spatial autocorrelation and over-estimate significance. 2 is the ideal case which is rarely realised in practice, we know the data generating process so we can accurately sample from it. 3 is the new algorithm proposed in this paper which we hope will be comparable with method 2, especially in the region where $I$ uniquely identifies $\beta$.

\begin{figure}[H]
    \centering
    \includegraphics[width=\textwidth]{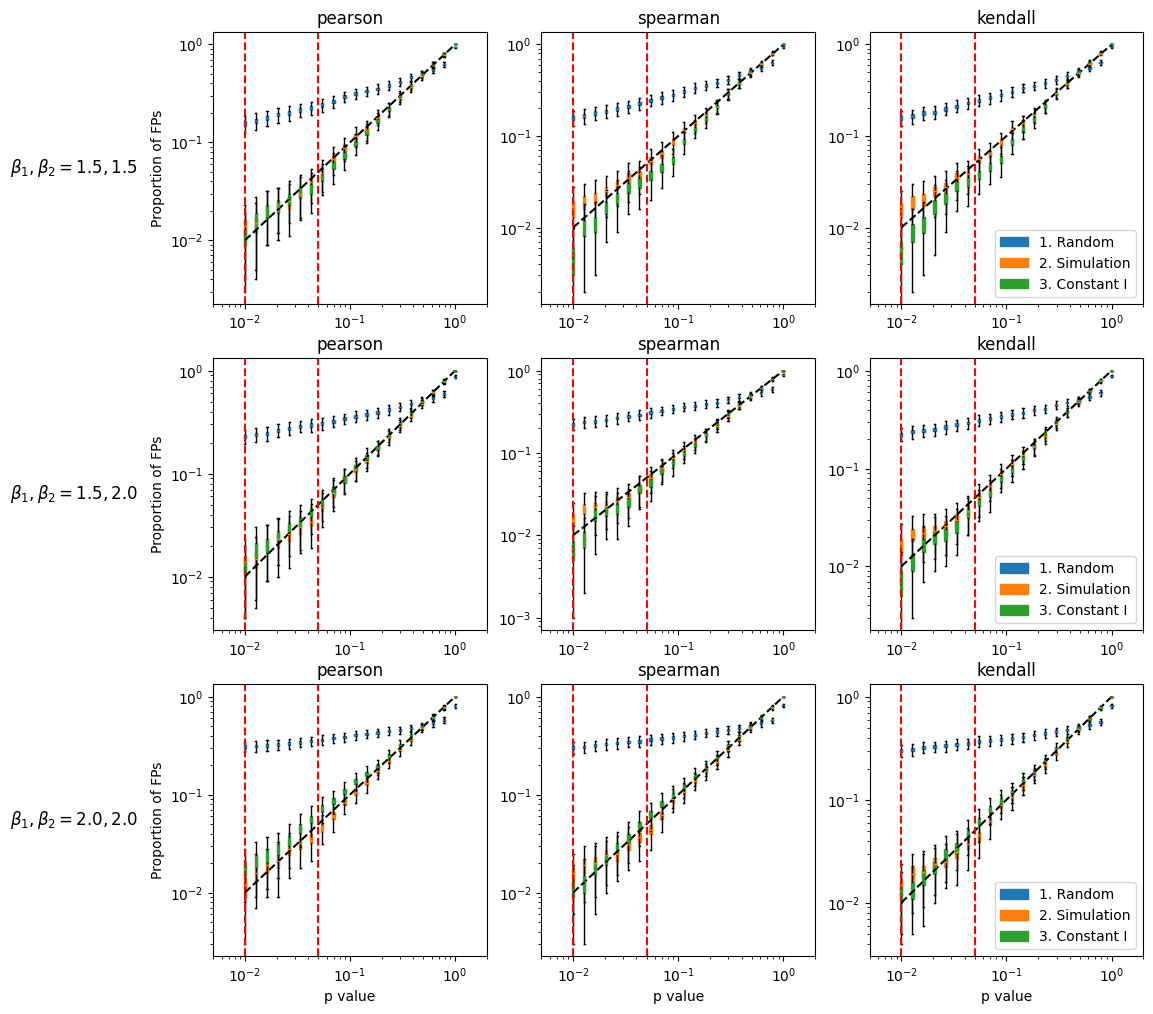}
    \caption{ An array of plots comparing the p-values estimated using each of the 3 sampling approaches described in the text to the observed number of false positives. The ideal behaviour is indicated by the black diagonal line. Vertical red lines indicate 0.01 and 0.05 significance levels. Each row corresponds to a different pair of $(\beta_1, \beta_2)$ values and each column to a different statistic. Note that the orange and green boxes overlap completely in most cases.}
    \label{fig:fptest}
\end{figure}

Figure \ref{fig:fptest} shows the expected versus observed number of false positives for three pairs of $(\beta_x, \beta_y)$ for the Pearson, Spearman and Kendall correlation statistics. To create this we generated 1000 pairs of independent random fields, $x_1$ and $x_2$, then for each of these 1000 pairs used each of the methods 1-3 to estimate the p-value for the corresponding statistic. We use $N_p=100$, which gives $N_p(N_p-1)/2 = 4900$ values to estimate the distribution of each statistic. The proportion of false positives is computed by counting the proportion of those 1000 pairs which would be said to be significantly correlated at each significance level i.e. there should be around $50$ significantly correlated pairs in $1000$ experiments at a significance level of $0.05$. Variance, indicated by the boxes, is estimated using bootstrap resampling.

We see that 1. random permutations consistently generate a high number of false positives, as expected this is not a reliable method to asses significance when there is spatial autocorrelation. Data generated by the same process that generated the `observed' data, method 2, results in accurate p-values and the expected number of false positives at each significance level. Our algorithm, 3, produces very similar results to 2, the green and orange boxes almost completely overlap most of the time in Figure \ref{fig:fptest}, with some slight deviations at lower p-values where variance is expected to be higher. Based on these results we propose that resampling configurations at fixed $I$ will give comparable results to simulating data from the real generating process, and hence allow accurate estimation of statistical significance in cases where this is not known.

\subsection{Caveats}

\begin{figure}[H]
    \centering
    \includegraphics[width=0.8\textwidth]{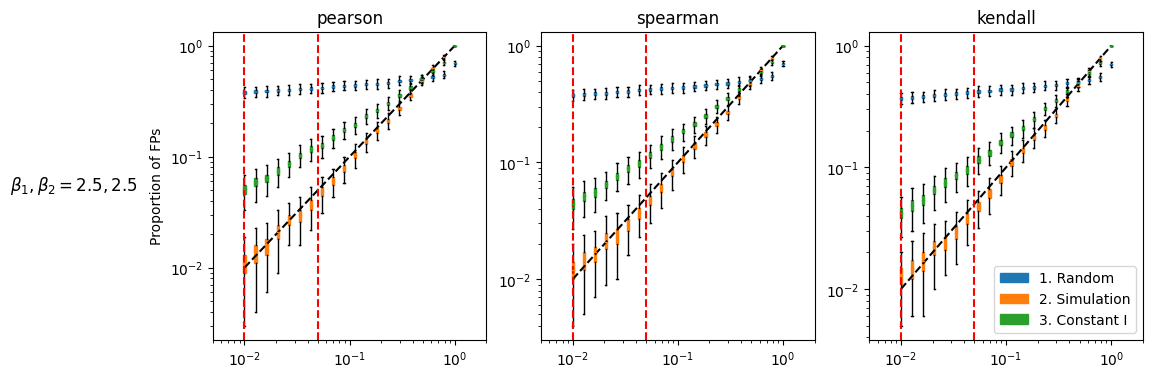}
    \caption{ An array of plots comparing the p-values estimated using each of the 3 sampling approaches described in the text to the observed number of false positives. The ideal behaviour is indicated by the diagonal back dashed line. Vertical red lined indicate 0.01 and 0.05 significance levels. Showing that for $(\beta_1, \beta_2) = (2.5,2.5)$ the algorithm of this paper overestimates the number of false positives. }
    \label{fig:testhi}
\end{figure}

While the results of the previous section support our method, we offer some caveats in this section. We do not expect that a single, fairly simple, statistic like Moran's $I$ can characterise every possible data generating process. For example, Figure \ref{fig:moranvbeta} shows that $I$ cannot distinguish values of $\beta$ larger than $\sim 2.5$, the values bunch up, so that an observation with $I \sim 0.8$ could plausibly be generated by 
$\beta$ between about $2.5$ and $3$. Figure \ref{fig:testhi} shows how, for larger $\beta$, and hence $I$, our method yields too many positive results. By tuning certain algorithm parameters (see Appendix \ref{sec:app}) it is possible to get the constant $I$ method `line up', with the correct number of false positive predictions. However, above $I \sim 0.7$ the algorithm starts to become sensitive to the parameters used for the optimisation and we don't recommend it.

We also note that some statistics (for example, the Kolmogorov-Smirnov statistic $sup_s |F_x(s) - F_y(s)|$) are more sensitive than others to differences in the real and constant $I$ distributions, especially when using the resampling approach discussed in Appendix \ref{sec:app}. Investigating, and potentially overcoming these limitations will be the focus of future work, for now we recommend our method in cases where autocorrelation is not excessively high, $I \lessapprox 0.7$, and for statistics, like correlation, which sum over all the data points.

\section{Guerry Data}\label{sec:guerry}

\begin{table}[H]
    \centering
    \begin{tabular}{|c|c|c|c|c|}
\hline
Literacy &Desertion &Commerce &Donation Clergy &Clergy \\ \hline
0.718 &0.630 &0.514 &0.428 &0.421 \\ \hline
\end{tabular}
    \caption{Moran Index for 5 Guerry variables.}
    \label{tab:moran_tab}
\end{table}

\begin{table}[H]
    \centering
\begin{tabular}{|c|p{0.1\textwidth}|p{0.1\textwidth}|p{0.1\textwidth}|p{0.1\textwidth}|p{0.1\textwidth}|}
\hline
&Literacy &Desertion &Commerce &Donation Clergy &Clergy \\ \hline
Literacy &&1) \textit{0.000}\newline 3) 0.106 &1) \textit{0.000}\newline 3) \textbf{0.017} &1) \textit{0.000}\newline 3) 0.097 &1) 0.068\newline 3) 0.316 \\ \hline
Desertion &&&1) \textit{0.001}\newline 3) 0.116 &1) \textit{0.000}\newline 3) 0.089 &1) \textit{0.021}\newline 3) 0.248 \\ \hline
Commerce &&&&1) 0.080\newline 3) 0.286 &1) 0.225\newline 3) 0.379 \\ \hline
Donation Clergy &&&&&1) \textit{0.001}\newline 3) \textbf{0.046} \\ \hline
    \end{tabular}
    \caption{Pearson correlation between 5 Guerry variables. P-values less than 0.001 are truncated to zero. 1 is the random permutation method, 3 is the method of this paper, resampling at constant $I$. `Significant' results, p-values less than $0.05$, according to method 1 are indicated by italics and significant results as judged by method 3 are in bold.}
    \label{tab:guerry_corr}
\end{table}

\begin{figure}[H]
    \centering
    \includegraphics[width=\textwidth]{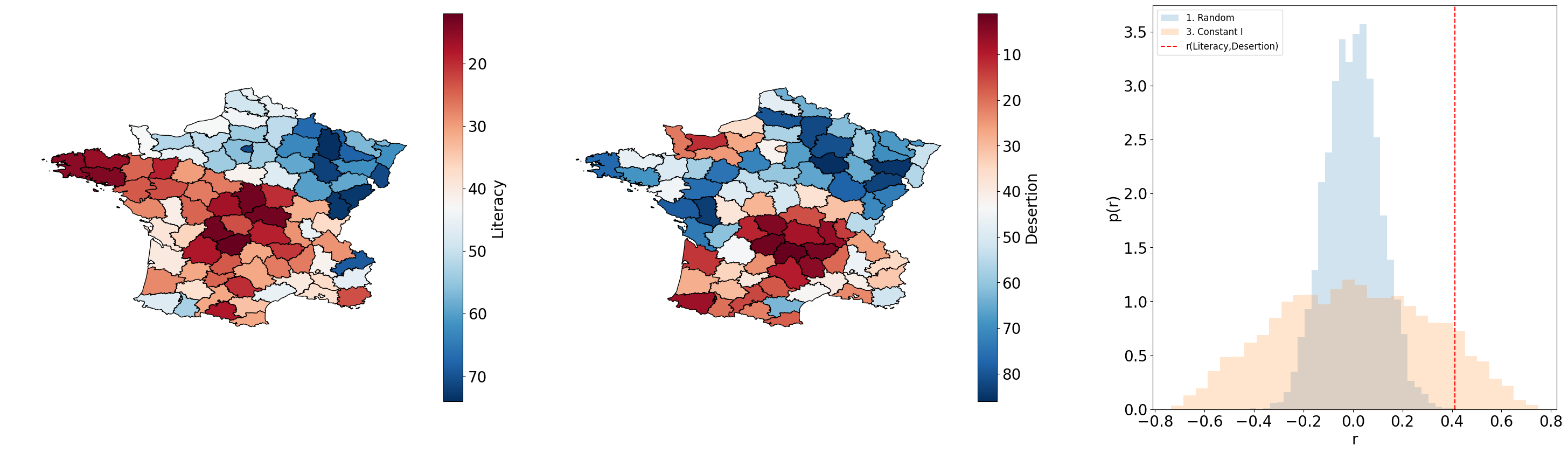}
    \caption{ The Guerry Literacy and Desertion statistics. Both are strongly spatially autocorrelated. The observed Pearson correlation between the variables is 0.41, but this is only moderately high compared to other possible arrangements of this data that have similarly large autocorrelation. }
    \label{fig:guerrydist}
\end{figure}

For a test on real data we return to the Guerry dataset. We choose 5 variables with high spatial autocorrelation, see Table \ref{tab:moran_tab}, and compute p-values for Pearson correlation (the analysis for the other correlation metrics is much the same) in Table \ref{tab:guerry_corr}. The table shows that the random permutation approach indicates a large number of significant correlations (6 at significance level $\alpha=0.05$) while our approach, taking spatial correlations into account, finds fewer, 2 at significance level $\alpha=0.05$ and none at significance level $\alpha=0.01$. This does not mean that, for example, the maps for Literacy and Desertion, are not similar, see Figure \ref{fig:guerrydist}. Rather it means that this level of correlation is expected between data with this degree of spatial autocorrelation. In this sense, autocorrelation alone `explains' the observed similarity between the variables. This does not rule out every correlation, the correlation of Literacy with Commerce and Clergy with Donations to the Clergy is still significant. The data presented in Figure \ref{fig:guerry}, Wealth versus Lottery Wager, are also significantly correlated under this null model ($r = 0.493$, $p = 0.002$).

\begin{figure}[H]
    \centering
    \includegraphics[width=\textwidth]{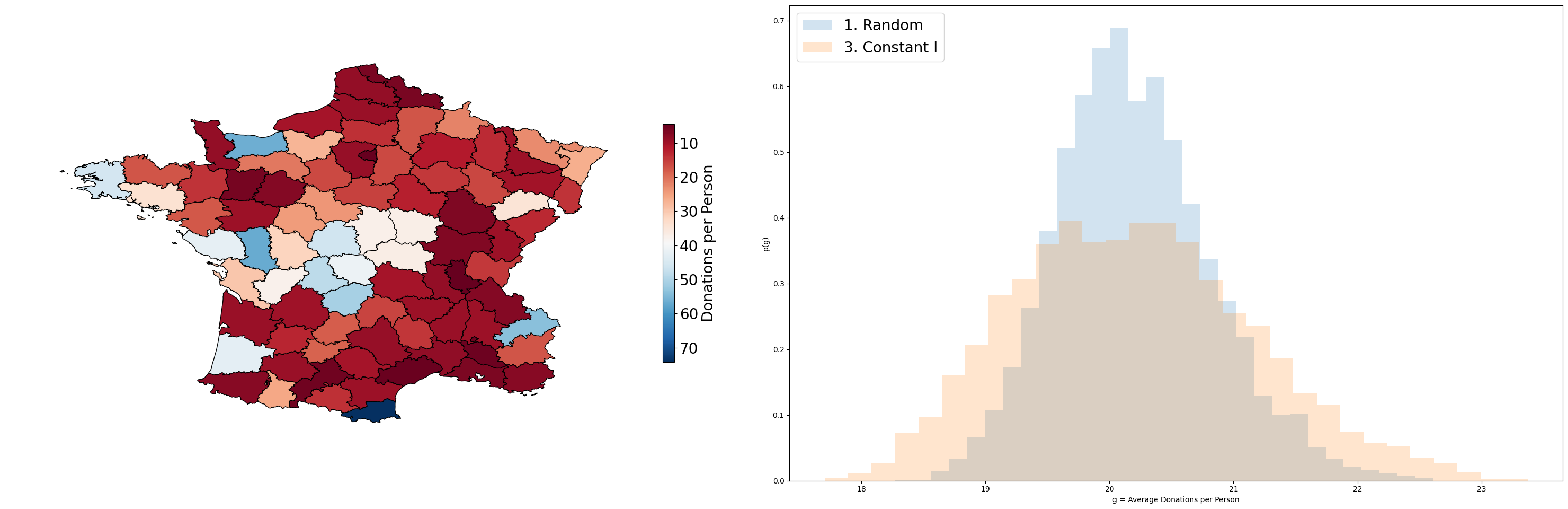}
    \caption{ The Guerry Donations divided by the Pop1831 data. The mean value per sampling area is $19.6$. The distribution on the right estimates the variance of this mean, with and without control for spatial autocorrelation. }
    \label{fig:guerrygenerosity}
\end{figure}
This method can also be used to estimate variances. As an example, from the Guerry data we can measure the mean Donations per Population. We can then resample the Donations and Population data to obtain the distribution shown in Figure \ref{fig:guerrygenerosity}. In this case both distributions are similar, with the Constant $I$ sampling estimating a larger standard deviation compared to Random Sampling: $19.60 \pm 0.95$ versus $19.60 \pm 0.62$. If the degree of spatial autocorrelation is fixed, then we are sampling configurations which are more likely to have `clusters' of high and low values, which leads to higher variance. In contrast, there is a very small probability to see such clustered configurations in an ensemble of completely random permutations, giving a lower variance.

\section{Discussion}\label{sec:conc}

In this paper we have introduced a simple but powerful method for significance testing in the presence of spatial autocorrelation - resample the data while keeping the autocorrelation fixed. Using Moran's $I$ to measure spatial autocorrelation, we introduced an algorithm to perform this resampling in Section \ref{sec:alg} (and discussed in more detail in Appendix \ref{sec:app}). We showed that the method performs excellently in cases where the outcome is known. It gives similar results to the recommendations of \cite{lennon2000red} and \cite{dale2002spatial} to generate samples using the same process that generated the data. The crucial advantage of this method is that \emph{we do not have to model the data generating process}. Finally we showed two examples of how the method may be used in practice in Section \ref{sec:guerry}, for significance testing and estimation of variance.

While we believe the method described here is immediately practical there are a number of possible ways to develop it. Other autocorrelation measures, like Geary's C , Getis–Ord G \citep{cliff1981spatial}, Lee's spatial smoothing scalar \citep{lee2001developing}, the Hopkins statistic \citep{hopkins1954new} etc. could be studied and may have superior properties to Moran's I. We note that our approach struggles at very high levels of spatial autocorrelation and perhaps other statistics would fare better, this is an open question left for future work. While the methods developed here are fast enough for analysing the small to medium datasets considered, for larger datasets other optimisation algorithms could also be considered.

Comparing spatial patterns is a very common problem. Such comparisons are frequently made across many disciplines. To give only a few examples: in sociological data analysis, crime versus poverty \citep{lockwood2007mapping}; in remote sensing, satellite imagery versus physical measurements \citep{nanzad2019ndvi} or the many ecological examples highlighted by \citep{lennon2000red}. We believe that this method is sufficiently general and simple enough to be used across disciplines to rigorously compare spatial patterns and we hope that it finds wide use among practitioners of spatial data analysis. To aid in this, we make our implementation freely available as a python package: \href{https://pypi.org/project/spatialsignificance/}{https://pypi.org/project/spatialsignificance/}.

\appendix

\section{Greedy Optimisation}\label{sec:app}

\begin{algorithm}
\caption{Greedy Optimisation of the Moran index}\label{alg:one}
\KwData{$\tilde{x}$ is a random permutation of the data $x$}
\KwResult{$x_p$ such that $|I_{target} - I(x_p)| \leq  \epsilon$}
\While{$|I_{target} - I(\tilde{x})| > \epsilon$}{
  1. Propose an update $\tilde{x} \gets \tilde{x}'$\\
  2. Compute $dI = I(\tilde{x}') - I(\tilde{x})$\\
  $dE \gets dI( 2(I_{target} - I(\tilde{x})) - dI)$\\
  \If{$dE \geq 0$}{
    $\tilde{x} \gets \tilde{x}'$
  }
}
$x_p \gets \tilde{x}$
\end{algorithm}

As discussed in Section \ref{sec:alg}, we generate configurations of similar Moran index to the observed data using a greedy optimisation approach. Starting with a random permutation of the data, $\tilde{x}$, the target function
\begin{equation}
E = (I_{target} - I(\tilde{x}))^2
\end{equation}
is minimised using algorithm \ref{alg:one}. One could equally well use a finite temperature approach like simulated annealing \citep{kirkpatrick1983optimization} or any other optimisation algorithm. The above is found to work well in practice. For steps 1 and 2 we give two possible approaches, permutation and resampling.

\subsection{Permutation}
This approach consists of swapping values of $\tilde{x}$ at randomly chosen positions $a$ and $b$. Using equation \ref{eqn:moran2} we can see that the term under the square root is fixed under permutations, only the $r(\tilde{x}, \tilde{x}l)$ term changes. Moreover, the denominator in Equation \ref{eqn:pearson} is also unchanged under permutations so only the change in the numerator needs to be computed. Updates consist of swaps $\tilde{x}_a \leftrightarrow \tilde{x}_b$ where $a \neq b$, which induce changes: $\tilde{x} \rightarrow \tilde{x}'$,  $\tilde{x}l \rightarrow \tilde{x}l'$ and hence $r(\tilde{x}, \tilde{x}l) \rightarrow r(\tilde{x}', \tilde{x}l')$. Defining $z_i = (\tilde{x}'_i - \bar{\tilde{x}})$ after some simple but tedious algebra we get that
\begin{equation*}
\Delta r = r(\tilde{x}', \tilde{x}l') - r(\tilde{x}, \tilde{x}l) \propto (z_a - z_b) ( {zr}_b - {zr}_a + {zl}_b - {zl}_a - (z_a - z_b)( w_{ab} + w_{ba} ) ) 
\end{equation*}
The proportionality constant is one over the denominator in equation \ref{eqn:pearson}. This formula can be used in algorithm \ref{alg:one} to compute $dI$ and hence $dE$. If a swap is accepted, we update the lagged vectors using
\begin{equation*}
{zl} (a \leftrightarrow b)_i = {zl}_i - w_{ia} (z_a-z_b) + w_{ib} (z_a-z_b) 
\end{equation*}
with a similar update scheme for $zr$. Note only indices $i$ that are neighbours of $a$ and $b$ need to be updated.

\subsection{Resampling}
This approach consists of replacing the value of $\tilde{x}$ at randomly chosen positions $a$ with values $\tilde{x}'_a$ sampled at random from the observed data. The corresponding update formulas are again the result of straightforward algebra, and are somewhat more complicated than the permutation approach. 
\begin{align*}
d &= (\tilde{x}'_a - \tilde{x}_a) \\ 
\Sigma_{lr} &= \sum_{j} (\tilde{x}l_j + \tilde{x}r_j)\\
\Delta \Sigma_{lr} &= d \sum_{j} (w_{aj}  + w_{ja} ) \\
\Delta r_1 &= d \sum_{j} (w_{aj}  + w_{ja} )x_j \\
\Delta r_2 &= (\bar{\tilde{x}} + d/N)(\Sigma_{lr} + d \Delta \Sigma_{lr} ) - \bar{\tilde{x}}(\Sigma_{lr} )\\
\Delta r_3 &= d \frac{|W|}{N} \left( 2 \bar{x} + \frac{d}{N} \right) \\
\Delta r &\propto \Delta r_1 - \Delta r_2 + \Delta r_3
\end{align*}
The denominator in equation \ref{eqn:moran} also changes,
\begin{align*}
SS &= \sum_i (\tilde{x}_i - \bar{\tilde{x}})\\
\Delta SS_1 &= (x'_a)^2 - (x_a)^2 \\
\Delta SS_2 &= d ( 2 \bar{\tilde{x}} + d/N) \\
\Delta SS &= \Delta SS_1 - \Delta SS_2
\end{align*}
Using $I' = \frac{N}{|W|} \frac{r(\tilde{x}, \tilde{x}l) + \Delta r}{SS + \Delta SS}$ we can calculate $dI$ and hence perform algorithm \ref{alg:one}. As above, if a swap is accepted, to save recomputing from scratch we update the lagged vectors using
\begin{equation*}
\tilde{x}l (a \leftrightarrow b)_i = \tilde{x}l_i + d w_{ia}
\end{equation*}
with a similar update scheme for $\tilde{x}r$. We  also update the sum $\Sigma_{lr} \rightarrow \Sigma_{lr} + \Delta \Sigma_{lr}$ and $\bar{\tilde{x}} \rightarrow \bar{\tilde{x}} + d/N$ to save recomputing those. The extra complcation of the formula is compensated by fewer iterations of algorithm \ref{alg:one} required before convergence.

\subsection{Starting Configuration}

\begin{algorithm}
\caption{Greedy Optimisation of $I$}\label{alg:two}
\KwData{$\tilde{x}$ is a random permutation of the data}
\KwResult{$x_p$ such that $I(x_p)$ is $>2I_{target}$ or is maximised to tolerance $\eta$}
$I_{old} \gets I(\tilde{x})$\;
$c \gets 0$\;
\While{($c \% N = 0$ and $I(\tilde{x}) - I_{old} > \eta$) or ($I(\tilde{x}) < 2 I_{target}$)}{
  1. Propose an update $\tilde{x} \gets \tilde{x}'$\\
  2. Compute $dI = I(\tilde{x}') - I(\tilde{x})$\\
  \If{$dI \geq 0$}{
    $\tilde{x} \gets \tilde{x}'$\;
  }
  $c \gets c+1$
}
$x_p \gets \tilde{x}$
\end{algorithm}

\begin{figure}[H]
    \centering
    \includegraphics[width=\textwidth]{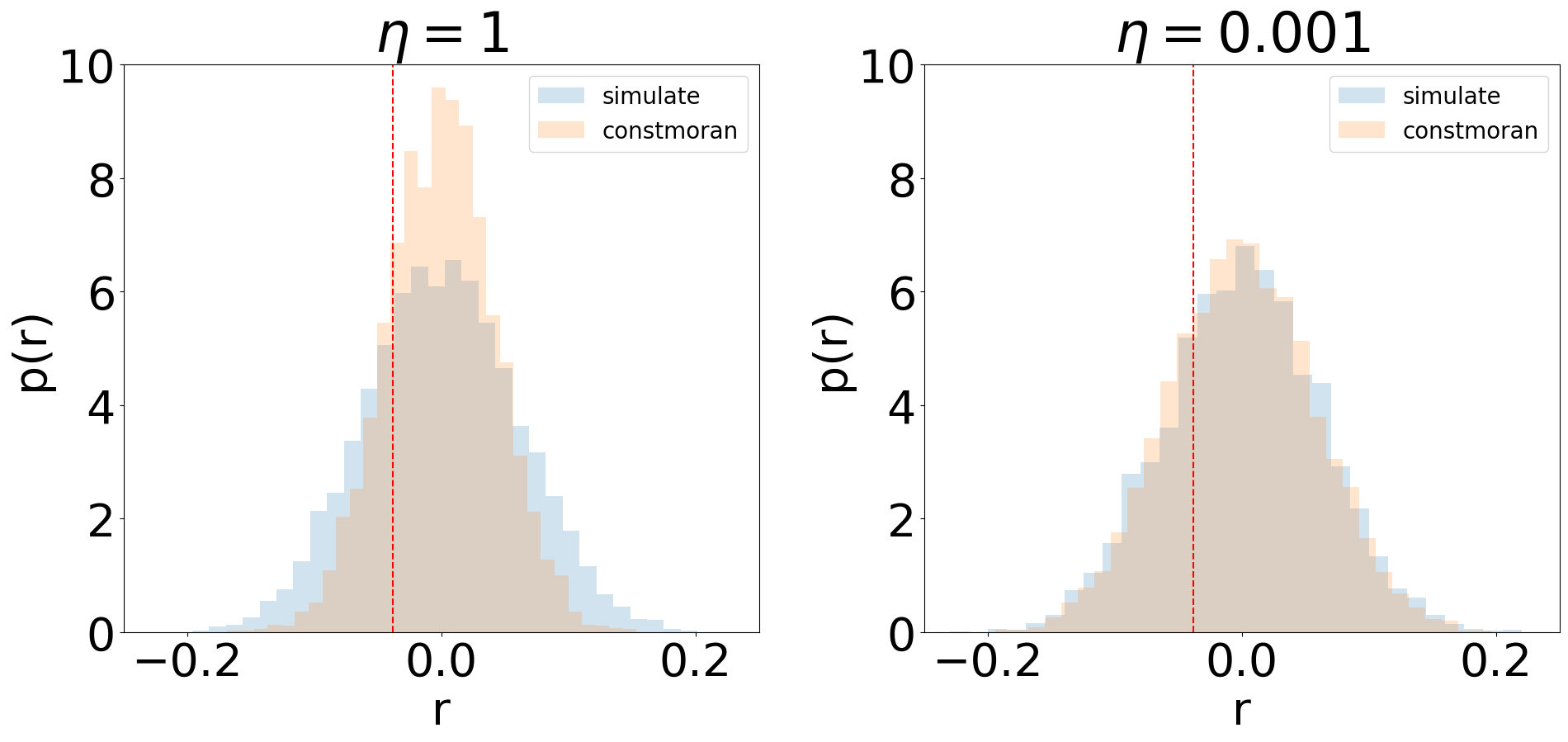}
    \caption{Left: Estimating the distribution of Pearson correlation $r$ using algorithm \ref{alg:one}. Right: Estimating the distribution of Pearson correlation $r$ using algorithm \ref{alg:two} then \ref{alg:one}. }
    \label{fig:prefreeze}
\end{figure}

We observe in practice that we cannot achieve the correct false positive ratio on our test set using algorithm \ref{alg:one} unless we approach the target Moran index from above rather than below i.e. it is better to start with a highly autocorrelated configuration and add noise than the induce order in a noisy configuration. To do this in practice we  first run algorithm \ref{alg:two}, then use the output as the starting configuration for algorithm \ref{alg:one}. Algorithm \ref{alg:two} performs a greedy optimisation of $I$ (using a permutation or resampling approach), stopping when either a configuration with double the target Moran index is reached or, when this is not possible, (i.e. if $I>0.5$) when $N$ updates have been proposed which don't increase $I$ by more than some tolerance $\eta$. Figure \ref{fig:prefreeze} shows a comparison of the obtained distribution of Pearson correlation scores, with and without using the `pre-freezing', algorithm \ref{alg:two}, compared to the correct distribution obtained by simulating new fields with the same value of $\beta$.

\begin{figure}[H]
    \centering
    \includegraphics[width=\textwidth]{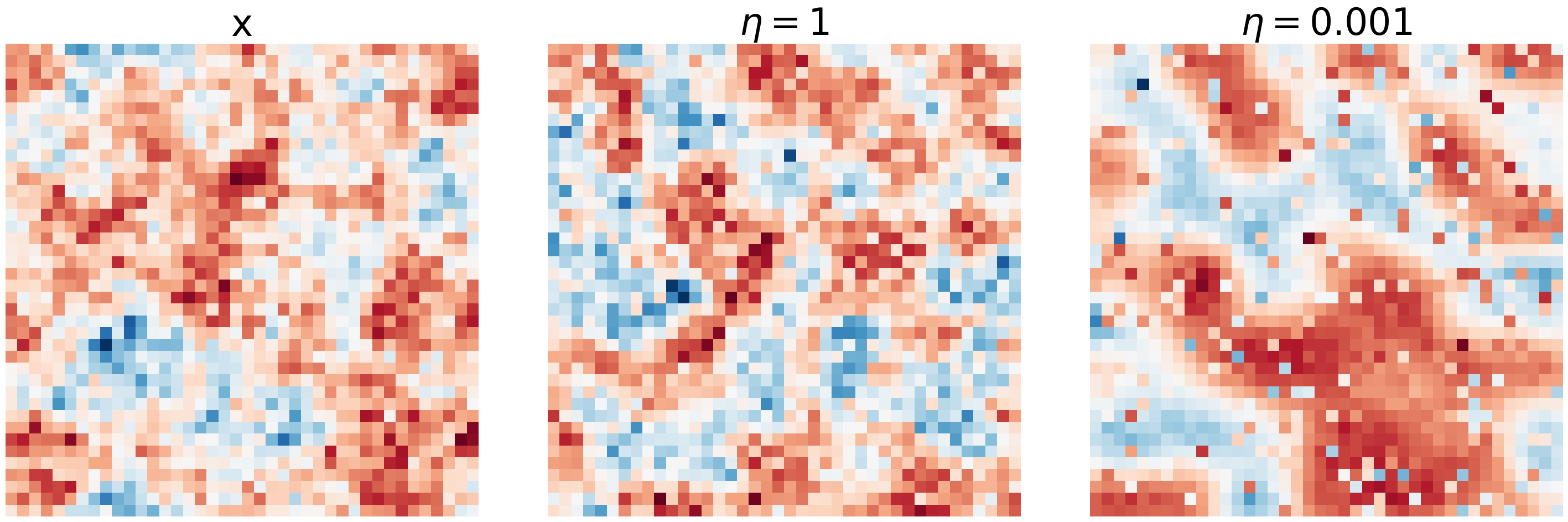}
    \caption{Left: The observed data generated with $\beta = 2$. Middle: A sample with the same $I$ computed with algorithm \ref{alg:one}. Right: A sample with the same $I$ computed using algorithm \ref{alg:two} then \ref{alg:one}. }
    \label{fig:freezecomp}
\end{figure}

A heuristic explanation for this behaviour can be intuited from Figure \ref{fig:freezecomp}. Large, coherent `blobs' like the one on the right can potentially give very large (or large and negative) correlations but are very unlikely to be generated by algorithm \ref{alg:one} starting from a random state. If these states are not sampled, the tails of the distribution will be too thin. Ideally, one would sample configurations using a multi-canonical \cite{berg1992multicanonical}  or density of states based approach \cite{wang2001efficient} so that the whole configuration space is covered. Preliminary experiments indicate that this is possible but extremely slow, however further work is required. In practice the pre-freezing approach works well up to $I \sim 0.7$ with some dependence on the value of $\eta$ at higher $I$.

\subsection{Algorithm Parameters}

\begin{figure}[H]
    \centering
    \includegraphics[width=0.8\textwidth]{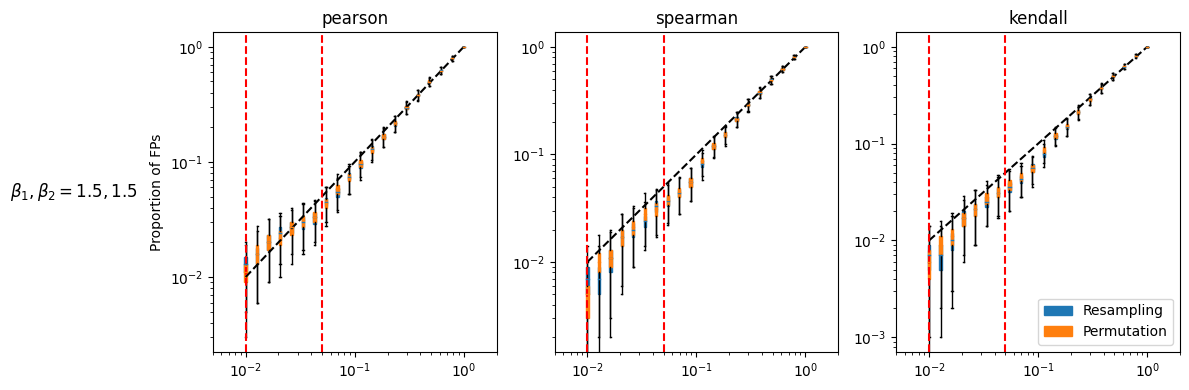}
    \caption{ Validation plot, constructed in the same way as Figure \ref{fig:fptest}, only showing the algorithm described in this paper comparing the permutation and resampling approach. }
    \label{fig:testopt}
\end{figure}

\begin{figure}[H]
    \centering
    \includegraphics[width=0.8\textwidth]{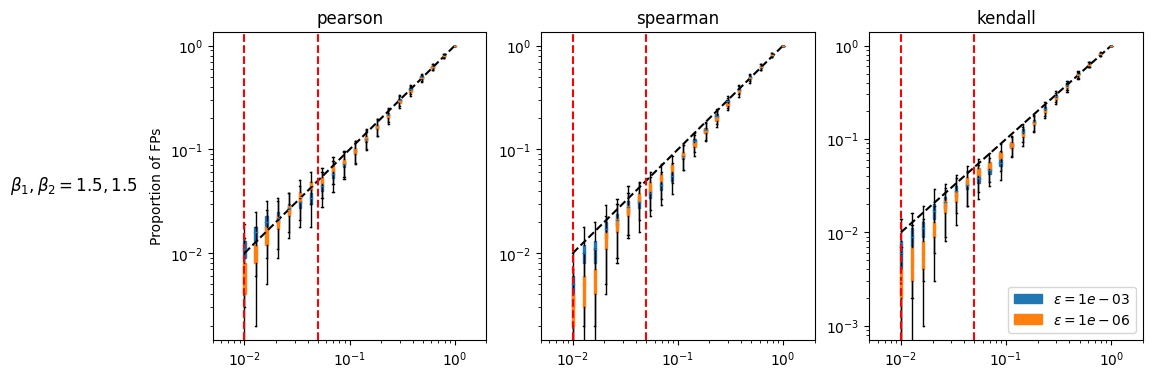}
    \caption{ Validation plot, constructed in the same way as Figure \ref{fig:fptest}, only showing the algorithm described in this paper for different values of the convergence parameter $\epsilon$. }
    \label{fig:testeps}
\end{figure}

Figure \ref{fig:testopt} shows a similar plot to Figure \ref{fig:fptest}, comparing optimisation using permutation and resampling. The results are quite similar, with the resampling approach converging in fewer iterations, though each iteration is more costly so the total time is comparable. Figure \ref{fig:testeps} shows a similar plot to Figure \ref{fig:fptest} using different values of $\epsilon$, there is a slight reduction in variance but substantial overlap so the outcome does not depend sensitively on $\epsilon$.

\bibliographystyle{plainnat}
\bibliography{main}

\end{document}